\documentstyle[preprint,aps]{revtex}
\begin{document}
%
%
\def\sqr#1#2{{\vcenter{\vbox{\hrule height.#2pt
        \hbox{\vrule width.#2pt height#1pt \kern#1pt
          \vrule width.#2pt}
       \hrule height.#2pt}}}}
\def\square{\mathchoice\sqr34\sqr34\sqr{2.1}3\sqr{1.5}3}
%
%
\title{Quantum manifestations of classical chaos in a Fermi accelerating disk}
\author{R. Badrinarayanan$^{1}$, Jorge V. Jos\'{e}$^{1,2}$ and G. Chu$^1$\\
{\it
$^1$Department of Physics, Northeastern University,\\ Boston
Massachusetts 02115, USA\\ and\\
$^2$Instituut voor Theoretische Fysica,\\ Princetonplein 5,
Postbus 80006, 3508 TA Utrecht, The Netherlands\\}}
\maketitle
\begin{abstract}

We study the classical and quantum mechanics of a
two-dimensional version of a Fermi accelerator. The model consists
of a free particle that collides elastically with the walls of a circular
disk with the radius varying periodically in time. A complete quantum
mechanical
solution of the problem is possible for a specific choice of the time-periodic
oscillating radius. The quasi-energy spectral properties of the model
are obtained from direct  evaluation of finite-dimensional
approximations to the time evolution operator. As the scaled $\hbar$  is
changed from large to small the statistics of the Quasienergy Eigenvalues
($QEE$) change from Poisson to circular orthogonal ensemble ($COE$).
Different statistical tests are used to characterize this transition.
The transition of the  Quasienergy Eigenfunctions ($QEF$) is also studied
using the $\chi ^2$ test with $\nu$ degrees of freedom. The Porter-Thomas
distribution is shown to apply in the $COE$ regime, while the Poisson regime
does not fit the $\chi ^2$ test with $\nu=0$. We find that the Poisson
regime is associated with exponentially localized $QEF$
whereas the eigenfunctions  are extended in the $COE$ regime.
To make a direct comparison between the classical and quantum solutions
we change the representation of the model to  one
in which the boundary is {\it fixed} and the Hamiltonian acquires a
quadratic term with a time-periodic frequency. We then carry out a
successful comparison between specific classical phase space surface-of-section
solutions and their corresponding quasi-energy eigenfunctions in
the Husimi representation.

\end{abstract}


\section{INTRODUCTION}
\label{sec:intro}

In recent years a good deal of attention has been directed at trying to
understand the  quantum manifestations of classical chaos (QMCC).
Although a complete understanding is not yet in sight,
significant progress has been made in obtaining partial answers
to this paradigm. This progress has primarily been achieved from studies of
lower-dimensional models: 2-D for energy conserving models and
1-D for driven systems. In particular, very few studies have been
carried out in two-dimensional time-dependent problems, for even in the
classical limit the theoretical analysis is nontrivial. It is the purpose
of this paper to consider a time-periodic two-dimensional problem which,
because of the  particular nature of the model, has a  reduced degree of
complexity and thus allows us to analyze its solutions in detail.

The model considered here is a two-dimensional version of the thoroughly
studied one-dimensional Fermi accelerator \cite{ll83}.
The 1-D quantum problem was introduced in Ref. \cite{jc89}. There it was
shown that for a very  specific form of the wall-oscillation function the
Floquet evolution operator can be explicitly
written and analyzed, without the complications of numerical
time-ordering (which effectively precludes a thorough investigation).
The model has been further analyzed by others  \cite{bill,chu92,seba}.
The Fermi acceleration disk (FAD) studied here consists  of a free particle
bouncing elastically inside a circular boundary with a radius that oscillates
periodically in time.
Classically, the model exhibits a transition to chaos as the amplitude of
the wall oscillation is increased. In the quantum case, the periodicity in the
wall oscillations allows us to use
Floquet's theory in terms of the one-period time evolution operator.

This model is of interest for several reasons: first, because the dynamics
is not kicked, as in almost all other time-periodic
models considered, {\it e.g.} the thoroughly investigated periodically kicked
rigid rotator model (PKR) \cite{i90}. Second, the explicit form of the
tight-binding-like model obtained for the dynamical equations
decays algebraically instead of exponentially, as in the PKR.
The FAD model then allows for a comparison of our results to those
of the kicked models, for it is important to know if the corresponding results
are generic. Third, the possibility
exists here that the spectrum of the evolution operator contains a continuous
component \cite{seba},
implying  nonrecurrent behavior of the wave functions.
Evidence suggestive of such a continuous spectrum has been reported in
\cite{chu92}.

One of the major advances in the understanding of the QMCC comes from
the clear differences found between the eigenvalue spectral properties of
models that classically exhibit chaos and those that do not \cite{bgs84}.
The tools used to measure these differences are extensions of those developed
in Random Matrix Theory (RMT) \cite{m91}.
In this paper we present an analysis of the
spectral properties of the time evolution operator using these tools.

While the FAD model provides considerable evidence for the spectral (or
statistical) characterization of the transition from integrability to
chaos, for reasons explained later, it turns out to be inconvenient in the
study of the classical-quantum correspondence in phase-space. To overcome
this difficulty, we map the FAD model onto a model of
a particle subjected to a bounded, rotationally symmetric, inverted harmonic
oscillator potential in the presence of time periodic kicks (
the Fermi inverted parametric oscillator or FIPO model)\cite{chu92}.
In this new representation, we {\it can} construct a mapping in terms of
classical canonical variables, which in turn permits us to make
explicit connections between classical
phase space solutions and their corresponding quantum counterparts
in terms of the Husimi distributions. This change in representation also
allows us to make connections with experiments in mesoscopic quantum dots
\cite{bj94}.

The outline of the paper is as follows: In section II we define both the
classical and quantum FAD and FIPO models.
The quantum problem is shown to be almost completely
integrable for the specific functional form of the oscillating radius.
In Section  III we present the bulk of our quantum results for the
statistics of the $QEE$ in the FAD representation, both for eigenvalues and
eigenfunctions.
Specifically, we discuss the eigenvalue statistical properties in terms
of different measures that include the nearest-neighbor spacing probability
distribution and its integrated form, the $\Delta_3$ statistic and
the two-point function $\Sigma _2$. In fitting the data from the
Poisson to the $COE$ regime we use the Brody function as well as the Izrailev
tests to try and quantitatively parametrize the transition region.
 We further discuss the eigenfunctions properties  in different
regimes using the $\chi ^2$ distribution of $\nu$ freedoms as a convenient
parameterization of the results\cite{al86}, as well as
their localization properties in terms  of the
time-averaged transition probabilities. In Section IV we discuss
the explicit classical-quantum correspondences in phase space, carried out
in the FIPO representation. We consider a series of classical
orbits in the surface of section and then identify the
Husimi contour plots of the eigenfunctions of the Floquet evolution operator
that correspond to the classical solutions. Finally, in section V we present
both our conclusions and also some questions left for future studies.

\section{THE CLASSICAL AND QUANTUM MODELS}
\label{sec:fad}

As we mentioned in the introduction we will use two representations of
the model studied here. Both representations have their own strengths
and thus their analysis leads to complementary results.
In this section we explicitly define the FAD and FIPO models in the classical
and quantum limits. The classical mechanics of the FAD is discussed
briefly while the classical solutions of the FIPO are discussed in more
detail since it will be in the FIPO representation that
the quantum-classical correspondence will be carried out.

\subsection{Classical Fermi accelerating disk (FAD)}
\label{subsec:cfad}

The FAD model consists of a free particle confined to move inside a
two-dimensional disk whose radius oscillates periodically in time;
$R(\tau)=R_0\theta(\tau)$ with $\theta(\tau+T)=\theta(\tau)$ and $T$
is the period. The  circular wall is taken to represent an
infinite potential barrier for the motion of the particle, that is,
classically the particle  undergoes  perfectly elastic collisions,
while quantum mechanically
the particle's wavefunction vanishes identically at the boundary.

The classical Hamiltonian  written in polar coordinates is
\begin{equation}
H = {1\over 2m}( p_r^2\>+\>p_{\phi}^2 ),
\end{equation}
where $r$ is the radius with $0\leq r\leq R(\tau)$, $\phi$ the azimuthal angle,
and $m$ the mass of the particle. The resulting equations of motion are,
\begin{mathletters}
\label{all1}
\begin{equation}
\label{eq:1a}
{{d^2r\over d\tau^2} = {J^2\over m^2r^3}}, \qquad
0\leq r(\tau) \leq R_0\,\theta(\tau),
\end{equation}
\begin{equation}
\label{eq:1b}
{{d\phi\over d\tau} = {J\over mr^2}}, \qquad
0\leq \phi(\tau) < {2\pi},
\end{equation}
\end{mathletters}
where $J$ is the magnitude of the {\it conserved} orbital angular momentum.
In principle, we can study the dynamics of the problem
for any form of the wall oscillation
function $\theta(\tau)$. However, as we shall show in the quantum case,
the problem dictates the appropriate $\theta(\tau)$  that allows an
explicit evaluation of the time evolution operator. The specific form of
$\theta (\tau)$ is found to be \cite{jc89,chu92}
\begin{equation}
\label{eq:theta}
\theta(\tau) = \sqrt{1+2\epsilon\left|{\rm mod}(\tau,{T\over 2})\right|},
\end{equation}
where $\epsilon$ is the amplitude of the wall oscillation. We will henceforth
measure $\epsilon$ in units of $R_0$ and
time in units of the period $T$, while the mass will
be taken as one henceforth.  Note that
$\theta(\tau)$ is already discontinuous in its first derivative, which
is important in order to have chaotic solutions.

As mentioned above, the FAD representation is not the most appropriate
one to discuss the classical-quantum correspondence of the solutions,
of central interest in this paper. However, to show that the FAD model
does indeed show a transition to chaos we have solved the classical equations
of motion at a period for different parameter values.
Examples of  $p_r$ vs $r$ phase space  plots are shown in Figs. 1(a-c)
for three values of $J$ and $\epsilon$.
Note that since $\phi$ is related to $r(\tau)$ by Eq.\ (\ref{eq:1b}),
if $r(\tau)$ shows irregular behavior, so  will $\phi (\tau)$.

\subsection{Classical finite inverted parametric oscillator (FIPO)}
\label{sec:cfipo}

To obtain the FIPO representation of the model studied here we begin by
carrying
out the Liouville transformation\cite{bellman},
\begin{eqnarray}
\label{eq:loui}
\tau &= \int_0^t \theta^2(t')\,dt' ,\\
r(\tau) &= \theta(t)\,\rho(t).
\end{eqnarray}
resulting in the new equation of motion for the $\rho$ coordinate,
\begin{equation}
{d^2\rho \over dt^2}\>+\> \omega^2(t)\,\rho\>-\>{J^2\over m^2\rho^3}=0,
\> \quad 0 \leq \rho(t) \leq R_0,
\end{equation}
where
\begin{equation}
\label{eq:7}
\omega^2(t) = {d\over dt}\left({1\over \theta}{d\theta\over dt}\right)\>-\>
\left({1\over \theta}{d\theta\over dt}\right)^2.
\end{equation}
The equation of motion can be derived from  the transformed Hamiltonian
\begin{equation}
\label{eq:8}
{H_{FIPO} = {p^2\over 2m}\>+\>{1\over 2}m\,
\omega^2(t)\,\rho^2\>+\>{J^2\over 2m\rho^2}},
\> \quad {\rm{with}}\quad 0 \leq \rho \leq R_0.
\end{equation}
This hamiltonian describes the one-dimensional motion of a particle in an
oscillator potential whose frequency changes with
time, as given by equation Eq.(\ref{eq:7}), with  a centrifugal barrier
and confined to a {\it stationary} infinite well of width $R_0$.
If, in addition, we choose the particular form of the wall oscillation function
$\theta(t)$ as given in equation (\ref{eq:theta}), the explicit time-dependent
frequency of the oscillator is given by
\begin{equation}
\omega^2(t) = -\,\epsilon^2\>+\>2\,\epsilon \sum_{n=-\infty}^{\infty}
\{\delta(t\,-\,(n+{1\over 2})\,T_0)\>-\>\delta(t\,-\,nT_0)\},
\end{equation}
where
\begin{equation}
T_0 = \int_0^T {d\tau\over\theta^2(\tau)} = {1\over \epsilon}\,
\ln(1+\epsilon T).
\end{equation}
Thus, the oscillator frequency consists of a constant, {\it negative} part,
and two sets of periodic Dirac $\delta$-functions, one at integer, and the
other at half-integer periods. This is the same form for $\omega(t)$
obtained by Chu and Jos\'e in their study the one-dimensional Fermi
model \cite{chu92}.

In normalized units, the full classical Hamiltonian $H_{FIPO}$ reads
\begin{equation}
H_{FIPO} = {p^2\over 2}\>-\>{1\over2}\,\epsilon^2\,\rho^2\>+\>
{J^2\over 2\rho^2}
+\>2\,\epsilon\,\rho^2\sum_{n=-\infty}^{\infty}\{\delta(t-(n+{1\over 2}))
-\delta(t-n)\}\>.
\end{equation}
The free parameters in the problem are $\epsilon$ (associated with the
amplitude of wall motion) and $J$ (the angular momentum of the particle).
Note that the harmonic oscillator kick strength alternates in sign
every half-period.

When the particle is free, that is when it does not hit the wall  nor does
it get kicked, the equation of motion is
\begin{equation}
\label{eq:9}
\ddot \rho = \epsilon^2\,\rho + {J^2\over\rho^3}.
\end{equation}
The solution to Eq.(\ref{eq:9}) is easily verified to be
\begin{eqnarray}
\label{eq:10}
\rho(t) &= \sqrt{\left[\rho_0\,\cosh\{\epsilon (t-t_0)\}+
\>{\strut \displaystyle{p_0}\over\displaystyle{\epsilon}}\,
\sinh \left\{\epsilon (t-t_0)\right\} \right]^2\>+\>{\strut\displaystyle{J^2}
\over
\displaystyle{\rho_0^2\epsilon^2}}\,\sinh^2 \left\{\epsilon (t-t_0)\right\}
},\\
 & p(t) = \dot\rho(t),
\end{eqnarray}
or, in terms of the initial energy
\begin{equation}
\label{eq:10a}
E_0 = {1\over 2}\,\left(p_0^2\>-\>\epsilon^2\,\rho_0^2\>+\>{J^2\over
\rho_0^2}\right),
\end{equation}
the solutions are given by
\begin{eqnarray}
\label{eq:11}
\rho(t) &= \sqrt{a\,\cosh\left[ 2\epsilon(t-t_0)\>+\>\cosh^{-1}({\strut
\rho_0^2+E_0/\epsilon^2\over\displaystyle{a}})\right]\>
-\> {\strut\displaystyle{E_0}\over\displaystyle{\epsilon^2}} },\\
&p(t) = \pm\sqrt{2E_0\>+\>\epsilon^2\rho^2\>-\>{\strut\displaystyle{J^2}
\over\displaystyle{\rho^2}} }.
\end{eqnarray}
 The sign of $p$  is positive or negative
 depending on whether the particle's last collision was with the centrifugal
 barrier or the wall. Here, $t_0$, $\rho_0$ and $E_0$ are determined by the
initial conditions, and the constant $a$ is defined as
\begin{equation}
{a^2 = {J^2\over\epsilon^2}\>+\>{E_0^2\over\epsilon^4}}.
\end{equation}
For a given energy $E$ the radial coordinate is constrained by the
wall and the centrifugal barrier:
\begin{equation}
{\rho_{min}\leq\rho\leq 1\>,\qquad\quad
\rho_{min} = \sqrt{a\>-\>E_0/\epsilon^2}}.
\end{equation}
Note that these equations of motion are {\it non} linear, even in the wall's
absence. Formally, one can represent either of the set of equations
(\ref{eq:10})-(14) or (\ref{eq:11})-(17) as
\begin{equation}
\label{eq:12}
\left( \begin{array}{c}
\rho(t) \\
p(t)
\end{array}
\label{eq:12b}
\right) =
\left( \begin{array}{c}
F(\rho_0,p_0,t-t_0) \\
G(\rho_0,p_0,t-t_0)
\end{array} \right),
\end{equation}
but we can investigate the stability of the mapping by linearizing about fixed
points only numerically.

The effect of collisions with the wall (or potential barrier) is simply
to reverse the direction of motion of the particle :
\begin{equation}
\label{eq:13}
\left( \begin{array}{c}
\rho(t_c^+) \\
p(t_c^+)
\end{array} \right) =
\left( \begin{array}{cc}
1 & 0 \\
0 & -1
\end{array} \right)
\left( \begin{array}{c}
\rho(t_c^-) \\
p(t_c^-)
\end{array} \right),
\end{equation}
where $t_c$ is the time of collision of the particle with the wall
(or potential barrier), and the minus and plus signs indicate times
just before and after the collision.

The effect of the kicks at every half period are obtained
by integrating the equations of motion over an
infinitesimal kick time. The results are, for half-integer and integer period
kicks,
\begin{equation}
\label{eq:14}
\left( \begin{array}{c}
\rho(t_{{1\over 2}}^+) \\
p(t_{{1\over 2}}^+)
\end{array} \right) =
\left( \begin{array}{cc}
1 & 0 \\
{-2\epsilon} & 1
\end{array} \right)
\left( \begin{array}{c}
\rho(t_{{1\over 2}}^-) \\
p(t_{{1\over 2}}^-)
\end{array} \right),
\end{equation}
\begin{equation}
\label{eq:15}
\left( \begin{array}{c}
\rho(t_1^+) \\
p(t_1^+)
\end{array} \right) =
\left( \begin{array}{cc}
1 & 0 \\
{+2\epsilon} & 1
\end{array} \right)
\left( \begin{array}{c}
\rho(t_1^-) \\
p(t_1^-)
\end{array} \right),
\end{equation}
respectively. The non-integrability of the problem arises precisely
because of these kicks. In contrast to the one-dimensional (or $J=0$)
case, because the particle here can never be {\it exactly} at the origin
at finite energies, the kicks are always relevant (in the sense that they
always couple the particle's position and momentum).

While so far we have only looked at the $(\rho,p)$ plane, we can also take
a coordinate-space cut. From the definition of angular momentum,
\begin{equation}
{d\phi\over dt} = {J\over \rho^2(t)},
\end{equation}
we obtain that during a `free' propagation,
the angle advances according to
\begin{equation}
\phi(t)-\phi_0 = {1\over 2}\left\{ \sin^{-1}\left({1\over a}\left({1\over
\rho_0^2}-{E_0\over J^2}\right)\right) - \sin^{-1}\left({1\over a}\left(
{1\over \rho(t)^2}-{E_0\over J^2}\right)\right)\right\} \pmod {2\pi}
\end{equation}
where $a$ was previously defined. In this paper we concentrate on the
$(\rho,p)$ phase space plane.

We are now in a position to construct a map in phase space which takes an
initial set of coordinates to a final one. The algorithm actually used
was to propagate the initial conditions via Eqs. (\ref{eq:11})-(17)
until it either hit a barrier, in which case
Eq.(\ref{eq:13}) was used to change the phase space coordinates,
or until the time elapsed was a half-integer or integer multiple of the
period, in which case Eqs.(\ref{eq:14}) or (\ref{eq:15}) are used to transform
the coordinates. We emphasize that in between kicks there may be
many collisions
between the particle and the walls, which effectively means new initial
conditions for propagation under Eq. (\ref{eq:11})-(17).
In general, the map is quite complicated, and very sensitive to initial
conditions. By recording the values at each successive period, we obtain a
surface-of-section of the trajectory of the particle in phase space.

We shall discuss specific results under different conditions in Section IV.

\subsection{Quantum FAD model}
\label{subsec:qfad}

The Schr\"odinger equation of the quantum mechanical version of the FAD (QFAD)
model introduced in Section ({\ref{subsec:cfad}) in polar coordinates is
\begin{equation}
\label{scho}
i\,\hbar\,{\partial\Lambda\over \partial\tau} = -\,{\hbar^2\over 2m}\left\{
{\partial^2\over\partial r^2}\>+\>{1\over r}\,{\partial\over\partial r}
\>+\>{1\over r^2}\,{\partial^2\over\partial\phi^2}\right\}\,
\Lambda(r,\phi,\tau),
\end{equation}
with the boundary condition $\Lambda\left
(r(\tau)=R_0 \theta(\tau),\phi,\tau \right)=0$, and
normalization
\begin{equation}
\int\limits_0^{R_0 \theta(\tau)}r\,dr\int\limits_0^{2\pi}d\phi\, \left|
\Lambda(r,\phi,\tau)\right|^2  = 1.
\end{equation}
These equations completely specify the QFAD model considered
in this paper. To find the general  solution to the QFAD model
we expand the wave function $\Psi(r,\phi,\tau)$
in terms of the natural complete set of basis functions $\psi_{nl}$,
\begin{equation}
\label{eq:3}
\Lambda(r,\phi,\tau) = \sum_{l=-\infty}^{+\infty} \sum_{n=1}^{\infty}
C_{nl}(\tau)\,\psi_{nl}(r,\phi,\tau),
\end{equation}
where the $\psi_{nl}$'s are the
{\it instantaneous} orthonormalized eigenfunctions for a given $R(t)$,
{\it i.e.}\ the solutions to
\begin{equation}
\label{expansion}
-\,{\hbar^2\over 2m}\,\nabla^2\,\psi_{nl}(r,\phi,\tau) =
E_{nl}(\tau)\,\psi_{nl}(r,\phi,\tau).
\end{equation}
Here $\tau$ is a parameter and  $\psi_{nl}$ satisfies the same boundary
and normalization conditions as $\Lambda$. The  instantaneous solutions
are given by
\begin{equation}
\psi_{nl}(r,\phi,\tau) = {1\over{\sqrt\pi R(\tau)J_{l+1}(\beta_{nl})}}\>
J_l(\beta_{nl}\,{r\over{R(\tau)}})\> \exp(\,i \,\l\,\phi\,),
\end{equation}
where $J_l$ is the Bessel function of order $l$ and $\beta_{nl}$
is its n-th zero. The corresponding instantaneous energies are
\begin{equation}
\label{energy}
E_{nl}(\tau) = {\hbar^2\over 2m}\,\left({\beta_{nl}\over R_0\,\theta(\tau)}
\right)^2.
\end{equation}
We now substitute the expansion (\ref{eq:3})  into
the Schr\"odinger equation (\ref{scho}). After using orthogonality and
known identities involving Bessel functions, one can evaluate the integral
\begin{eqnarray}
K_l(m,n) &=\int\limits_0^1 x^2\,J_l({\beta_{ml}}x)\,{\strut\displaystyle{d}
\over\displaystyle{dx}}
J_l({\beta_{nl}}x)\,dx \nonumber\\
&={\strut \displaystyle{\beta_{ml}\beta_{nl}}\over\displaystyle{{\beta_{nl}^2 -
\beta_{ml}^2}}}\,
J_{l-1}(\beta_{ml})J_{l-1}(\beta_{nl}) \qquad\qquad(m\ne n) ,
\end{eqnarray}
so that the final result can be cast in the form
\begin{eqnarray}
\label{coef}
i\,\hbar\,\dot C_{nl} &=\sum_{m,k}\left[E_{nl}\delta_{k,l}\delta_{n,m}
 - i\,\hbar{\strut\displaystyle{\dot \theta}\over \displaystyle{\theta}}\,
\,{\strut \displaystyle{ 2\,\beta_{mk}\,\beta_{nk}}\over\displaystyle
{\beta_{mk}^2 - \beta_{nk}^2}}\delta _{k,l}(1-\delta_{m,n})\right]\,C_{mk} \\
 &\equiv \sum\limits_{m,k}\,H_{nm;kl}\,C_{mk} \nonumber,
\end{eqnarray}
where the overdot denotes differentiation with respect to $\tau$.
Equations (\ref{coef}) are  completely equivalent to the
Schr\"odinger equation (\ref{scho}).
The time development of the system is obtained from
\begin{equation}
C_{nl}(\tau) = \sum_{m,k}\,U_{nm;kl}(\tau,\tau ')\,C_{mk}(\tau ') ,
\end{equation}
with the time evolution operator defined by
\begin{equation}
U_{nm;kl}(\tau,\tau ') = {\bf T}\,{\rm exp}\left(-\,{i\over \hbar}
\int_{\tau '}^{\tau}
H_{nm;kl}(s)\,ds \right) ,
\end{equation}
where ${\bf T}$ formally represents the time-ordering operation. We can,
however, get rid of the time-ordering if the effective Hamiltonian
$H$ factors into an overall time-dependent factor multiplying a
time-independent part. For this to happen, we see from equations (\ref{energy})
and (\ref{coef}) that we must have
\begin{equation}
{{\dot \theta\over \theta} \sim {1\over \theta^2}} .
\end{equation}
An appropriate form for $\theta(\tau)$, satisfying the periodicity and
time reversal invariance conditions is the one given in (\ref{eq:theta}).
Using this form for $\theta (\tau)$ the matrix elements of the
one-period time evolution operator  are exactly given by
\begin{equation}
U_{nl;mk}(\tau+T,\tau) = {\rm exp}\left(-\,{i\over \hbar}\,H_{nl;mk}\,
\int_{\tau}^{\tau+T}
{ds\over \theta^2(s)} \right).
\end{equation}
Note that the  time evolution matrix is nominally four-dimensional but,
because of the conservation of angular momentum,
it reduces to a two-dimensional matrix, for a given value of $l$.

The required zeros of the Bessel functions are calculated
using the Newton-Raphson technique after
bracketing the root using the fact that
separation between successive zeros asymptotically approaches $\pi$
\cite{o60}. This method is  very efficient, since
finding the roots even for the high zeros and large $l$ values
can be done to machine accuracy.
The quasi-energy spectrum is determined once the zeros of the Bessel functions
are found and the $U$-matrix is
diagonalized numerically (with unitarity preserved to machine precision).
Since $U$ is unitary, its
eigenvalues are unimodular, and we write it's eigenvalue equation as
\begin{equation}
U(\tau+T,\tau)\,|{\lambda_n}\rangle = e^{i\varepsilon_n T}|{\lambda_n}\rangle
\qquad \quad {0} \leq \varepsilon_n \leq {2\pi}.
\end{equation}
Complete knowledge of $U$ implies {\it full} knowledge of the evolution
of the system at all integer multiples of the period because
$U(\tau+NT,\tau) = U^N(\tau+T,\tau)$. This product of  $U$'s
is the quantum counterpart of the classical one-period maps.
The quasienergy eigenvalues $\varepsilon_n$, being invariant under
unitary transformations,
are representation independent and the quasi-energy spectrum obtained this way
is of high quality from a numerical point of view. For our
calculations, we used a truncated basis of 400 states for calculating
properties of the $QEF$, while we used a basis of 500 states to
calculate the $QEE$. We checked that the $QEE$ spectrum was stationary,
{{\it i.e.} the eigenvalues remain basically unchanged to  within 1\% in this
case
as we increase the size of the $U$ matrix from 400$\times$400 to
600$\times$600.

\subsection{Quantum FIPO model}

Quantum mechanically, the required set of Liouville and unitary
transformations are a little
more complicated (because of the additional requirement of normalization),
but are nevertheless doable.
We first transform the Schr\"odinger equation (\ref{scho}) to an effectively
one-dimensional problem by writing the wavefunction as
\begin{equation}
\Lambda(r,\phi,\tau) = {1\over\sqrt{2\pi r}}\,e^{il\phi}\,\Phi(r,\tau).
\end{equation}
the Schr\"odinger equation (\ref{scho}) now reads
\begin{equation}
\label{eq:27}
i\,\hbar\,{\partial\over\partial\tau}\,\Phi(r,\tau) = -\,{\hbar^2\over 2m}
\,{\partial^2\over\partial r^2}\,\Phi(r,\tau)\>+\>{(l^2-{1\over 4})
\hbar^2 \over 2mr^2}\,\Phi(r,\tau) .
\end{equation}
We see from the above equation that the centrifugal potential barrier,
 $(l^2\,-\,{1\over 4})\,\hbar^2 / 2mr^2 $, diverges at the
origin for $l\!\neq\!0$.
We hereafter consider only the nontrivial case of nonzero angular momentum,
so we require in addition that $\Phi$ vanishes at the origin.
 The $\Phi$ normalization and boundary conditions are now
\begin{equation}
\label{eq:28}
\int_0^{R_0\theta(\tau)} |\Phi(r,\tau)|^2\,dr = 1 ,
\end{equation}
\begin{equation}
\label{eq:29}
\Phi(r=0,\tau) \>= 0 =\> \Phi(r=R_0\theta(\tau),\tau) .
\end{equation}
Equations (\ref{eq:27}-\ref{eq:29}) define an effective
one-dimensional problem for $r$.

We now proceed to apply the Liouville transformations, Eq.(\ref{eq:loui}).
the transformed wavefunction $\Psi(\rho,t)$ is obtained from the unitary
transformation \cite{chu92})
\begin{equation}
\label{eq:30}
\Phi(r,\tau) = {1\over\sqrt{\theta(t)}}{\exp}\left( i {m\over 2\hbar}
{\dot\theta\over\theta}\rho^2\right)\Psi(\rho,t),
\end{equation}
where the overdot denotes now differentiation with respect to $t$.
Following the definitions and applying the rules of partial
differentiation, we have
\begin{eqnarray}
{\partial\over\partial r}& =& {\partial\rho\over\partial r}\,{\partial\over
\partial\rho}+{\partial\over\partial r}{\partial\over\partial t} =
{1\over\theta(t)}{\partial\over\partial\rho} \\
 {\partial^2\over\partial r^2} &=& {1\over\theta^2(t)}\,{\partial^2\over
\partial\rho^2} \\
{\partial\over\partial\tau} &=& {\partial\rho\over\partial\tau}{\partial
\over\partial\rho}+{\partial t\over\partial\tau}{\partial\over
\partial t} = -{\rho\dot\theta\over\theta^3(t)}{\partial\over\partial
\rho}+{1\over\theta^2(t)}{\partial\over\partial t}.
\end{eqnarray}
Applying these results to the set of
equations (\ref{eq:27}-\ref{eq:30}), after some algebra, we
finally obtain Schr\"odinger equation satisfied by $\Psi(\rho,t)$,
\begin{equation}
\label{eq:31}
i\,\hbar\,{\partial\over\partial t}\,\Psi(\rho,t) = \left\{-\,
{\hbar^2\over 2m}\,
{\partial^2\over\partial\rho^2}\>+\>{1\over 2}\,m\omega^2(
t)\rho^2\>+\>{(l^2-{1\over 4})\hbar^2\over 2m\rho^2}\,\right\}\,
\Psi(\rho,t) ,
\end{equation}
along with the transformed normalization and boundary conditions
\begin{equation}
\int_0^{R_0} |\Psi(\rho,t)|^2\,d\rho = 1 ,
\end{equation}
\begin{equation}
\Psi(\rho=0,t)= 0 =\Psi(\rho=R_0,t).
\end{equation}
The corresponding quantum FIPO (QFIPO) Hamiltonian is
\begin{equation}
H_{QFIPO} = {p^2\over 2m}+{1\over 2}m\omega^2(t)
\rho^2+{J^2\over 2m\rho^2} ,
\end{equation}
where
\begin{equation}
J \equiv \hbar\sqrt{l^2-1/4} \qquad l > 0 .
\end{equation}
The hamiltonian $H_{QFIPO}$ is simply the quantized version of the
corresponding classical Hamiltonian. What is significant is that the
frequency law for $\omega^2(t)$
is the same in both classical and quantum mechanics.
As pointed out earlier in \cite{chu92}, this is a special effect arising
out of the classical action consisting only of linear and quadratic terms.

The one-period time evolution operator is then given (in units where the mass,
time and length are scaled in units of $m$, $T_0$ and $R_0$ respectively)
by
\begin{equation}
\label{eq:32}
U(1,0) = \exp(i{\epsilon\over\hbar}\rho^2)\,\cdot\,\exp(-i{H_0
\over 2\hbar})\,\cdot\,\exp(-i{\epsilon\over\hbar}\rho^2)\,\cdot\,
\exp(-i{H_0\over 2\hbar}) ,
\end{equation}
where
\begin{equation}
\label{eq:33}
H_0 = {p^2\over 2}\>-\>{1\over 2}\,\epsilon^2 \rho^2\>+\>
{J^2\over 2\rho^2} .
\end{equation}
The first and third terms in Eq.(\ref{eq:32}) correspond
to the kicks at full and
half periods, while the other two terms correspond to the evolution of the
state under $H_0$. We note here that while
the classical problem has two free parameters $\epsilon$ and $J$, in the
quantum
case we have the additional parameter $\hbar$. By writing the
Schr\"odinger equation (\ref{eq:31}) in terms of dimensionless quantities, it
is easy to verify that the parameter dependence of $U$ is via $\epsilon$,
$\epsilon/\hbar$ and $J$.

For convenience we choose a set of Fourier sine functions
as the basis on which to diagonalize the evolution operator
to span the Hilbert space. While unitarity requires a large number of
Fourier components as the scaled $\hbar$ gets smaller, this choice of basis has
several advantages -it not only ensures automatic satisfaction of the boundary
conditions, but also  considerably simplifies the numerical calculations
involved. In a sense, the Fourier basis is also the natural choice to
handle the very short-wavelength oscillations encountered in the QFAD model.
The details of obtaining the matrix elements of the evolution operator in
this basis are presented in the Appendix.

\section{STATISTICS OF $QEE$ SPECTRA IN THE QFAD MODEL}
\label{subsubsec:sqfad}

As mentioned in the introduction, one of the clear QMCC
emerges when one compares the spectral properties  of specific model systems
as appropriate parameters are tuned to classically produce a transition
from integrable to completely chaotic regimes. In this section we follow
the general thinking developed in RMT to implement different
tests to quantify the spectral properties of the QFAD model.
These properties are obtained from  a direct diagonalization
of the one-period time evolution matrix. For the results presented here
we vary the value of $J$ and $\epsilon/\hbar$, while
we take the wall amplitude fixed at $\epsilon=1$. This value corresponds to
a non-perturbative value, where the kinetic energy, the
centrifugal barrier and the energy imparted by the wall all have
about equal magnitudes. Furthermore,
we rescale all lengths by $\epsilon$ and all energies by $\hbar$.
The semiclassical limit of interest here is then obtained when
$l\uparrow\infty,\>\hbar\downarrow 0$, with $
J=\hbar\sqrt{l^2-{1\over 4}}$
kept constant.

Next we discuss the RMT tests and their application to the results obtained
for the $QEE$ of the QFAD model.

\subsection{Nearest neighbor $QEE$ distributions}

A local measure often used in RMT is the distribution of nearest-neighbor
energy level separations, $P(s)$, where $s=\varepsilon_{n+1}-\varepsilon_n$.
In the extreme integrable and chaotic regimes it has been established
that $P(s)$ takes the Poisson or Wigner distribution forms,
\begin{eqnarray}
P_P(s) =   e^{-s} ,\\
P_W(s) &=  {\pi\over 2}\,s\,e^{-{\pi\over 4}s^2},
\end{eqnarray}
respectively.
A convenient and often successful parameterization of the $P(s)$ obtained
in the transition between $P_P$ to $P_W$ is provided
by the Brody interpolation formula \cite{b81}:
\begin{equation}
P_{\alpha}(s) = \gamma(\alpha+1)\,s^{\alpha}\,\exp(-\gamma s^{\alpha+1}),
\end{equation}
where $\gamma = \left[\Gamma\left({\alpha+2\over\alpha+1}\right)\right]^
{\alpha+1}$, and $\Gamma(x)$ is the Gamma function. This distribution is
normalized and, by construction, has mean spacing $\langle{s}\rangle=1$.
We recover the Poisson case taking $\alpha =0$ and Wigner for $\alpha =1$.
A criticism to the Brody distribution is, however, that there is no first
principles  justification for its validity. The fact remains that it
does fit the specific results found when considering explicit model systems.

Instead of first looking at the $P(s)$ we found that the fits are better if we
start by calculating the integrated distribution
\begin{equation}
\Pi _{\alpha}(s)=\int _{0}^{s}ds' P _{\alpha}(s'),
\end{equation}
where we use all the eigenvalue spacings calculated.
The results for $\Pi _{\alpha}(s)$ are shown in Fig. 2. Next we use the
Brody form to find the exponent $\alpha $ to fit the data to
$\Pi _{\alpha}=1-e^{-\gamma s^{\alpha +1}}$. The $\alpha$ exponent
is determined from a linear fit to $\ell n \left\{\ell n \left(1-\Pi _{\alpha}
\right)^{-1}\right\}= \ell n \{\gamma\} +(1+\alpha)\ell n\{s\}$.
The fits shown in Fig. 2 are obtained by using these values of $\alpha $.
These results  clearly show the
general trend. As the nonintegrability parameter $\epsilon/\hbar$
is increased, the $\Pi (s)$ goes from being Poisson-like to Wigner-like.
For small values of $s$ $(\leq 0.1)$, the Poisson limit is well fitted by the
Brody function while for the $COE$ the fit is not as good, as one would expect
due to the level repulsion.
The corresponding results for $P(s)$ are shown in Fig. 3, with the Brody
fits using the $\alpha $ values obtained from $\Pi _{\alpha}(s)$. We found that
the fits are better by analyzing the data this way rather than first
fitting $P(s)$ since in the binning process we lose information.

An alternative phenomenological interpolation formula with some justification
was proposed by Izrailev \cite{i89}
who used the analogy between the partition function of a two-dimensional
Coulomb gas of charged
particles on a circle (the Dyson gas) and the joint probability distribution
of the  quasi-energy eigenvalues of the $COE$
to  construct an approximate form for $P(s)$ given by
\begin{equation}
\label{eq:6}
P_{\beta}(s) = A \left({\pi\over 2}s\right)^{\beta} \exp \left\{
-{\beta\over 4}\left({\pi\over 2}s\right)^2 -
\left({2B\over \pi} - {\beta\over 2}\right){\pi\over 2}s
\right\},
\end{equation}
where $A(\beta)$ and $B(\beta)$ are constants fixed from
normalization and the condition $\langle{s}\rangle=1$. This formula also
reduces to the
appropriate Poisson and Wigner limits  for $\beta = 0$ and $1$, respectively.
As we discuss below, this interpolation formula is also good
for the QFAD but the Brody distribution provides a slightly better
fit to our results.
The interesting aspect of  Izrailev's distribution is that in principle
it provides a {\it quantitative} connection
between the energy level repulsion parameter $\beta$ and the degree of
localization of the $QEF$. The connection is established
by considering the ensemble averaged
``entropy localization length" $l_S$ for the $u_n$ components of
$QEF$ for a chosen basis and defined as
$\langle{l_S}\rangle = N \exp\left(\langle{{\cal S}_N}\rangle -
{\cal S}_N^{COE}\right),$
where the information entropy
\begin{equation}
{\cal S}_{N}(u_1,..,u_N) = - \sum_{n=1}^N w_n\,\ln\, w_n, \qquad  \qquad
w_n = \vert u_n \vert^2.
\end{equation}
Note that ${\cal S}_N$  is essentially the logarithm of the number of
sites significantly populated by the corresponding eigenstate, and
${\cal S}_N^{COE}$  is the entropy corresponding
to the random ($COE$) limit, introduced so that for completely chaotic states
the exact limiting value of $\langle{l_S}\rangle = N$. The conjecture is that
the quantity $\langle{l_S}\rangle/N$, which varies between 0 for completely
localized
states to 1 for fully extended  ones, is {\it identically} equal to the
repulsion parameter $\beta$ in Eq.(\ref{eq:6}).

To check the validity of this conjecture, we fitted the level spacing
distribution to $P_{\beta}(s)$ given in Eq.(\ref{eq:6}) and determined the
value $\beta_{hist}$ which minimized the $\chi^2$. We also calculated
the ensemble-averaged entropy
localization length as defined above, and calculated the corresponding
$\beta_{S}$, and compared the two. The results are presented in Figure 4.
We see that the agreement is good where the classical chaos is strong,
but gets worse as
the classical motion is more regular. Otherwise, the transition from $COE$ to
Poisson statistics is consistent with the general trends observed
previously. We conclude therefore that while the Izrailev model is
intuitively more appealing, the evidence to support it is not
compelling enough (in our model at least) to prefer it over the Brody formula.

\subsection{  $\Delta _3$ and $\Sigma ^2$ $QEE$ statistics}

We also calculated higher-order eigenvalue spectral correlations.
The average number of levels in an interval of length L is
$<n(L)> = {1\over L}\sum_{\alpha }n(\alpha ,L),$
where the $< >$ stands for spectral average, and $n(\alpha ,L)$ is the number
of
levels in an interval of length $L$ starting at $\alpha$ and ending at $\alpha
+
L$. Also important are the various moments of the level distribution. The one
considered here is the second moment of the average number of levels in
a given stretch of length $L$ of the spectrum, the $\Sigma^2(L)$
statistic \cite{bg84}
\begin{equation}
<\Sigma ^{2}(L)> = \left <(n(\alpha ,L) - <n(\alpha ,L)>)^2 \right >.
\end{equation}
Another often calculated statistic is the the Dyson-Mehta
$\Delta _3(L)$  which measures the stiffness of the  spectrum.
This is defined by
\begin{equation}
{\Delta _3(L,\alpha )}={1\over L}{min_{A,B}}{\int _{\alpha }^{\alpha +L}}
{[{\tilde N}(x)-Ax-B]^2}\,dx,
\end{equation}
where $\tilde N(x)$ is the unfolded number density. In our case there is no
need to unfold the spectrum. $\Delta _3$ is just the least mean square
deviation of $\tilde N(x)$ from the mean  straight line
behavior. This statistic is directly proportional to the
$<{\Sigma ^2}>$ by $\Delta _3 (L)={2\over {L^4}}\int _0^L(L^3-sL^2x+x^3)
\Sigma ^2(x)dx$, and thus can be calculated for the circular ensembles
as well. The specific theoretical predictions for the averaged
${<{\Delta_3(L)}>}={1\over L}{\sum _{\alpha }
\Delta_3(L,\alpha )}$, are ${{\Delta_3^{(COE)}(L)}}\>={{1\over \pi
^2}}\ell n\{L\} -0.007,\> \>$ and
${<{\Delta_3^{(Poisson)}(L)}>}\>={L\over 15}\> $.
These results are correct in the asymptotic limit valid for $15\leq L$.

In Fig. 5 we present our results for $<\Delta_3>$ and $<\Sigma ^2>$.
In these figures one clearly sees the transition
from Poisson-like (dashes) to $COE$-like (solid line) behavior
as $\epsilon/{\hbar}$ is varied. We note that the
$\Delta_3$ statistic does not saturate in the $COE$ limit, even for the maximum
interval $L$  that we looked at, as would be expected from semiclassical
arguments originally proposed by Berry. Furthermore, for the largest $L$
considered the Poisson limit does not present the knee seen in other completely
integrable systems. All in all the results shown in Fig. 5 are consistent
with what we have come to expect for the transition region.

\subsection{ $\chi^2$ eigenfunctions distribution}

To study the behavior of the eigenfunctions we begin by discussing
the statistics
of the overlap of the eigenfunctions with the natural basis vectors.
Several authors \cite{al86} have conjectured that as
the classical motion changes from chaotic to regular, this distribution
can be represented by a $\chi^2$-distribution in $\nu$ degrees of
freedom, with $\nu$ decreasing from 1
(the Porter-Thomas limit) to 0 (the Poisson limit):
\begin{equation}
P_{\nu}(y) = {(\nu /2)^{\nu /2}\over \Gamma (\nu /2)}\enskip y^{\nu /2 -1}
\enskip \exp(-\nu y/2),
\end{equation}
where $y \equiv \mid <\lambda |nl>\mid ^2$, $|{\lambda}\rangle$ label the
$QEF$ and $|{nl}\rangle$ label a set of N orthogonal basis vectors.
(The $y$'s have been  rescaled so that $\langle{y}\rangle=1$.) We have tested
this hypothesis for
the overlap strengths. The results, plotted on a logarithmic scale
in Figure 6, show the general trend of decreasing $\nu$ as the corresponding
classical system becomes more regular.
However, we note that as we move from the $COE$ to the Poisson
limit the fits to the
$\chi ^2$ get worse. Note especially
the appearance of a sharp second peak well away from 0 as $\nu$ decreases.
This discrepancy is related to the fact that the results are basis dependent
when not in the $COE$ limit. Equivalently, we do not expect to have a good fit
to the $\chi ^2$ except perhaps if we take the special basis obtained
from a semiclassical calculation. Even if we manage to get good agreement
with the $\chi ^2$ for a properly chosen basis
the result will not be generic and therefore the statistical analysis of
amplitudes would lose its universal meaning. Universality does apply, however,
in the $COE$ limit.

\subsection{Transition between localized to extended $QEF$}

The phenomenon of dynamic localization gives another QMCC\cite{casati}. In
this section we discuss the
localization properties of the quasi-energy eigenfunctions when projected
onto the natural basis. A physically meaningful quantity to calculate is the
time averaged transition probabilities $\overline {P(n,m)}$ where
$\{|{n}\rangle, n=1..N\}$ label the basis states.
While for small $\epsilon/\hbar$ the probabilities are
peaked around the diagonal ({\it i.e.}\, the $QEF$
are localized in energy space),
for larger values the probabilities are spread out, indicating delocalized
states. To quantify this transition
we define the  pair-correlation function $C(\Delta)$ by
\begin{equation}
C(\Delta) = {1\over N-\Delta}\sum_{m=1}^{N-\Delta} {\overline
P(m,m+\Delta)}\qquad
\Delta=0,1...N\!-\!1
\end{equation}
where ${\overline P(n_0,m)} = \ell im _{_T\rightarrow \infty}
{1\over {T}}\int _0^T\mid<n_0\mid m>\mid^2$ is the transition probability
from the initial unperturbed state $|{n_0l}\rangle$ to the
unperturbed state $|{ml}\rangle$. We find good fits of our evaluations
of this correlation by the form,
\begin{equation}
C(\Delta) \sim e^{-\Delta/\xi},
\end{equation}
where $\xi $ is a localization length.
The results are displayed in Fig. 7. We note that close to the
near-integrable regime $\xi << N$, indicating localization of the overlaps,
 while for larger $\epsilon/\hbar,\> \xi\geq N$. These results are consistent
with the statistical analyses and with the expectation of a smooth transition
from localized to extended solutions as we change from the Poisson to the
$COE$ regimes. We note that this results depends directly on th size of
the $U$ matrix being considered.

\section{CLASSICAL TO QUANTUM PHASE SPACE CORRESPONDENCE}

In this section we consider the question of how to relate the quantum
to the classical results. The natural habitat of nonintegrability
in hamiltonian systems is phase space. As we mentioned in the introduction
the FAD representation does not lend itself to an easy connection between
the classical and quantum results \cite{bill}. It is in the FIPO
representation that we can carry out this program. Here we start by discussing
a set of explicit classical phase space FIPO solutions that we will then
directly connect to the QFIPO model results. The connection is done by
using the Husimi representation of the quasi-energy eigenfunctions.

\subsection{Classical FIPO model results}

In Figs. 8(a-c) we present results of the classical Poincar\`e sections for
three different angular momentum values, $l =1, 5, 10$, for the parameter
values indicated in the captions. We observe from these figures that there
is a symmetry line in the $(\rho,p)$ section of phase space.
As pointed out earlier\cite{chu92} for the one-dimensional case,
the reason for the existence of this symmetry line is as follows.
Consider a particle kicked at $t=0$. The position is unchanged, but the
momentum changes to $p^{(+)} = p^{(-)} + 2\epsilon\rho_0$. If we denote
$p^{(+)}$ by $p_0$, then at time $0^{(-)}$ the particle had momentum
$p^{(-)} = -2\epsilon\rho_0 + p_0$. Since the Hamiltonian is
time reversal invariant, we see that propagating a particle {\it forward}
in time from $(\rho_0,p_0)$ is the same as propagating it {\it backward}
from $(\rho_0,2\epsilon\rho_0-p_0)$. Thus, all results are symmetric about
the line $p = \epsilon \rho$. This symmetry is also present in
the quantum problem, as will be discussed below.

Although the map is complicated, there are a few cases which can be studied
analytically. The first of these is when the energy of the particle
arises solely from its angular motion. The particle then does not feel
the effects of the walls. The points $p_0 = \epsilon\rho_0$ lying
on the symmetry line are then fixed points of order one of the map.
To prove this,
we begin by noting that the energy corresponding to this initial condition
is given by Eq.(\ref{eq:10a}) is
\begin{equation}
\label{eq:16}
E_0 = {J^2\over 2\,\rho_0^2} ,
\end{equation}
while the momentum just after the kick at half period is
related to that just before by
\begin{equation}
\label{eq:17}
p_{{1\over 2}}^{(+)} = p_{{1\over 2}}^{(-)} - 2\,\epsilon\,\rho_{{1\over 2}} .
\end{equation}
The energy just after the kick is then related to that just before by
\begin{eqnarray}
\label{eq:18}
E_{{1\over 2}}^{(+)} &= {1\over 2}\left\{(p_{{1\over 2}}^{(-)} - 2\,\epsilon\,
\rho_{{1\over 2}})^2 - \epsilon^2\rho_{{1\over 2}}^2 + {J^2\over
\rho_{{1\over 2}}^2}\right\} \\
&= E_{{1\over 2}}^{(-)} + 2\,\epsilon\,\rho_{{1\over 2}}\,(\epsilon\,\rho_
{{1\over 2}} - p_{{1\over 2}}^{(-)} ) .
\end{eqnarray}
The trajectory of the particle during the second half-period is
simply the time reversed trajectory during the first half, so that
the energy has to be the same before and after the kick.
Thus, Eqs. (\ref{eq:17}) and (\ref{eq:18}) imply
\begin{equation}
\label{eq:19}
p_{{1\over 2}}^{(-)} = \epsilon\,\rho_{{1\over 2}}\quad , \quad
p_{{1\over 2}}^{(+)} = -\epsilon\,\rho_{{1\over 2}}.
\end{equation}
{}From Eq.(\ref{eq:19}), and using the definition of momentum from
Eq.(\ref{eq:12}),
\begin{equation}
p_{{1\over 2}}^{(-)} = \sqrt{ 2E_0 + \epsilon^2 \rho_{{1\over 2}}^2 -
{J^2\over \rho_{{1\over 2}}^2} } ,
\end{equation}
we can deduce that
\begin{equation}
\label{eq:20}
E_0 = {J^2\over 2\rho_{{1\over 2}}^2} .
\end{equation}
Comparing Eqs.(\ref{eq:16}) and (\ref{eq:20}) we see that
$\rho_{{1\over 2}} =
\rho_0$, and because of time reversal symmetry then tells us
that at $t=1$ we must have $\rho_1 = \rho_0$. The kick at full period
now changes the momentum
according to
\begin{equation}
{p_1^{(+)} = p_1^{(-)} + 2\,\epsilon\rho_0}.
\end{equation}
Using the same arguments for a full period as those used at half
period, we can see that
\begin{equation}
p_1^{(+)} = \epsilon\rho_0 = p_0 .
\end{equation}
Thus, $(\rho_0,\epsilon\rho_0)$ is a fixed point of order one of the map,
and numerically we can verify that it is of parabolic type, with
a maximum allowable value of $\rho_0$ for a given $\epsilon$ and $J$. That
is, this line of fixed points, which starts at $(\rho_{min},\epsilon
\rho_{min})$ ends at a point determined by the condition $\rho_{{1\over 2}}
\leq 1$. From substituting the initial conditions into Eq.(\ref{eq:10}) at
half-period, we find that this point is given by
\begin{equation}
\rho_0^{max} = {e^{-\epsilon/2}\over\sqrt{2}}\sqrt{ 1 + \sqrt{ 1 - \left(
{J\over\epsilon}\sinh({\epsilon\over 2})\exp({\epsilon\over 2})\right)^2 } }.
\end{equation}
The reality of this maximum imposes yet another restriction on the allowable
values of the parameters, {\it viz},
\begin{equation}
{{J\over\epsilon}\sinh({\epsilon\over 2})\exp(\epsilon/2) \leq 1}.
\end{equation}
We now proceed to calculate one of the simplest nontrivial period-2 orbits,
shown in Fig. 9. The particle starts at $(\rho_0,p_0)$ (shown as point
A in the figure), with energy $E_0$. The initial conditions are such that the
particle bounces off the wall at $\rho = 1$, reverses, hits the barrier at
$\rho = \rho_{min}$ (point C) and reverses momentum again before reaching
the point $(\rho_{{1\over 2}},p_{{1\over 2}}^{(-)})$ just before the kick
at half period (point D). Now, from Eq.{\ref{eq:11} we can write the time
elapsed in motion under the influence of the potential only as
\begin{equation}
\label{eq:21}
t-t_0 = {1\over 2\epsilon}\left\{ \cosh^{-1}\left({\rho^2+E_0/\epsilon^2\over
a}
\right) - \cosh^{-1}\left({\rho_0^2+E_0/\epsilon^2\over a}\right) \right\}.
\end{equation}
Using this result, and substituting the appropriate $\rho$'s, we can write down
the times elapsed during the three sections of the trajectory, $t_{AB}$,
$|t_{BC}|=t_{CB}$ and $t_{CD}$. Equating the sum of these three times to
1/2, we can determine the position at D in terms of $\rho_0$ and $E_0$.
The result is,
\begin{equation}
\label{eq:22}
\rho_{{1\over 2}} = \sqrt{ a\,\cosh\{ \epsilon + b(E_0) \}
- {E_0\over\epsilon^2} } ,
\end{equation}
where
\begin{equation}
\label{eq:23}
b(E_0) = 2\left\{\cosh^{-1}\left({\rho_{min}^2+{E_0/\epsilon^2}
\over a}\right) - \cosh^{-1}\left({1+{E_0/\epsilon^2}\over a}\right)\right\}
+ \cosh^{-1}\left({\rho_0^2+{E_0/\epsilon^2}\over a}\right).
\end{equation}
Now, the relation between the
momenta just before and after the kick, and the argument about
conservation of energy (based on time reversal invariance) lead to the
relation (\ref{eq:20}). Inserting this into Eq.(\ref{eq:22}), we finally get
\begin{equation}
\label{eq:24}
{J^2\over 2E_0} + {E_0\over\epsilon^2} = a\,\cosh\{ \epsilon + b(E_0)\} .
\end{equation}
Since $E_0$ is expressed in terms of $(\rho_0,p_0)$ and $a=a(E_0)$, the
above relation
provides, in principle, a relation between $\rho_0$ and $p_0$.

The particle, after the kick at $t=1/2$, follows the time reversed
trajectory to the point G (which is the same as A) at $t=1^-$ as shown,
at which point it will have position $\rho_0$ and momentum $-p_0$.
Then it gets kicked, which now changes its energy to $\tilde {E_0}$.
This time, the momentum after the kick is given by
\begin{equation}
p_1^{(+)} = -p_0 + 2\,\epsilon\,\rho_0 .
\end{equation}
Using conservation of energy under time reversal again, we find that
\begin{equation}
\tilde{E_0} = E_0 + 2\,\epsilon\,\rho_0\,(\epsilon\,\rho_0 - p_0) ,
\end{equation}
which also changes the other functions of energy:
\begin{equation}
\label{eq:25}
\tilde a = \sqrt{ (J/\epsilon)^2 + (\tilde{E_0}/\epsilon^2)^2 } ,
\end{equation}
\begin{equation}
{\tilde \rho}_{min} = \sqrt{ \tilde a - \tilde{E_0}/\epsilon^2 }.
\end{equation}
It is important to remember that both $\tilde a$ and ${\tilde \rho}
_{min}$ are functions of $\rho_0$ and $p_0$ via Eq.(\ref{eq:25}). After getting
kicked
at H, the particle hits the barrier at I, reverses and gets to point J
before getting kicked again, and following the time reversed trajectory
back to the initial point A, as shown. The logic of determining the
second relation between the initial conditions is the same as that used
during the first quarter of its trajectory. By setting the total time taken
$|t_{HI}|(=t_{IH})+t_{IJ}=1/2$, we first obtain a relation
similar to Eq.(\ref{eq:22}), and repeating the argument about
energy conservation due to time reversal invariance, we find
\begin{equation}
\tilde{E_0} = {J^2\over 2\rho_1^2}
\end{equation}
with
\begin{equation}
\rho_1 = \sqrt{\tilde a\,\cosh\left\{ \epsilon + c(\tilde{E_0})\right\}
- {\tilde{E_0}\over\epsilon^2}} ,
\end{equation}
and
\begin{equation}
c(\tilde{E_0}) = 2\cosh^{-1}\left({{\tilde \rho}_{min}^2+\tilde{E_0}/
\epsilon^2\over
\tilde a}\right) - \cosh^{-1}\left({\rho_0^2+\tilde{E_0}/
\epsilon^2\over \tilde a}
\right).
\end{equation}
Putting all these equations together, the final result is
\begin{equation}
\label{eq:26}
{J^2\over 2\tilde{E_0}} + {\tilde{E_0}\over\epsilon^2} = \tilde a\,\cosh\left\{
\epsilon + c(\tilde{E_0})\right\} .
\end{equation}
In principle, a physical solution to Eqs.(\ref{eq:24}) and (\ref{eq:26}) may be
found analytically for the fixed point $(\rho_0,p_0)$, in practice
 we find it numerically. The fixed point shown in the Fig. 8(b) (for $l=10$,
$\hbar=0.026$ and $\epsilon=1$), labelled F, is given by
$\rho_0 = 0.657385\ldots$,
$p_0 = 2.659181\ldots$. When substituted back, this
fixed point satisfies the equations given above to machine accuracy.

This analysis to determine this fixed point can be generalized to a family of
fixed points of period $2\cdot(m+n+1),\>\>$
with $n, m=0,1,\ldots$ by letting the particle bounce between
the wall and the centrifugal barrier $(2m+1)$ times during the first half
period (between B and C), and $(2n+1)$ times during the third (between H
and I). The equations to be solved are now
\begin{equation}
{J^2\over 2E_0} + {E_0\over\epsilon^2} = a\,\cosh\{ \epsilon + b_m(E_0)\} ,
\end{equation}
where
\begin{equation}
b_m(E_0) = (2m+1)\>b(E_0) - 2m\>\cosh^{-1}\left({\rho_0^2+
{E_0/\epsilon^2}\over a}\right) ,
\end{equation}
and,
\begin{equation}
{J^2\over 2\tilde{E_0}} + {\tilde{E_0}\over\epsilon^2} = \tilde a\,\cosh\{
\epsilon + c_n(\tilde{E_0})\} ,
\end{equation}
where
\begin{equation}
c_n(\tilde{E_0}) = (2n+1)\>c(\tilde{E_0}) - 2n\>\cosh^{-1}\left({{\tilde
\rho}_{min}^2+\tilde{E_0}/\epsilon^2\over \tilde a}\right) ,
\end{equation}
and $\tilde{E_0}$, $\tilde a$ and $\tilde\rho_{min}$ are obtained in analogy to
the period-2 case. In principle, higher order fixed points (and their
associated families) can be calculated analytically in the above manner by
writing larger numbers of such coupled nonlinear equations. In practice,
however, that is quite a formidable task. To determine the stable manifolds
we exploit the existence of the symmetry line around hyperbolic fixed points,
which are simply the unstable manifolds under time reversal.
In practice, numerical difficulties preclude determining the stable
manifolds in any other manner.

\subsection{Correspondence}

We can now make a direct comparison between the classical and quantum
FIPO results by employing a phase space approach. To do this, we
use the Husimi representation of the $QEF$\cite{chang}. The
Husimi distribution, interpreted as a probability density, is a
coarse-grained version of the Wigner function which goes smoothly to the
semiclassical limit. In practice, the most often used technique of
coarse-graining is to take the overlap of the $QEF$ with
coherent oscillator
states. For the radial coordinate the coherent state is
\begin{equation}
\Psi_{\rho_0,p_0}^G(\rho) = ({\sigma\over\pi\hbar})^{1\over 4}\,
\exp\left\{-{\sigma\over 2\hbar}(\rho - \rho_0)^2 + i{p_0\over \hbar}
(\rho - {\rho_0\over 2}) \right\},
\end{equation}
which is a minimum-uncertainty Gaussian wavepacket
centered at $(\rho_0,p_0)$, with root mean-squared deviations given by
$\Delta\rho = \sqrt{\hbar/2\sigma}$, $\Delta p = \sqrt{\hbar\sigma/2}$,
and $\sigma\,$ is the `squeezing' parameter. This parameter is adjusted
when making comparisons to the classical phase-space plots. The
Husimi distribution of a single $QEF$ $\psi_{\varepsilon}(\rho)$,
is then defined by
\begin{equation}
{\cal F}_{\varepsilon}(\rho_0,p_0) = \left| \>\int_0^1
\Psi_{\rho_0,p_0}^G(\rho)\,\psi_{\varepsilon}(\rho)\,d\rho\> \right|^2.
\end{equation}
The Husimi distribution is obtained by scanning through the values of
$(\rho_0,p_0)$ in the region of interest in phase space, and the result is
compared with the classical surface-of-section.
We begin the comparison by noting the symmetry about the line $p=\epsilon\rho$
in the Husimi contour plots in Figs. 10(a-d). As mentioned earlier, this
feature carries over from the classical results for the same reasons as there,
and it is in fact used to effectively halve the numerical effort.

All calculations reported here were carried out for the wall oscillation
amplitude $\epsilon=1$. In this case, all terms in the Hamiltonian
are comparable in magnitude, which means that we are in a
non-perturbative regime. A few calculations were done for different
$\epsilon$'s, but no new qualitative features emerged. In choosing a value of
$\hbar$, we were guided by the following considerations. The value of
$\hbar$ has to be small enough so that the system is well into the
semiclassical regime, yet large enough so that the dimension of the truncated
Hilbert space $N$ (which grows with decreasing $\hbar$) is large enough
to preserve
unitarity. Moreover, $N$ has to be such that the largest eigenenergy of
$H_{FIPO}$ has to be larger than the maximum energy of the classical
particle in the region of interest in phase space. These restrictions
eventually led us to choose $\hbar=0.026$ for $N=100$. {\it All} the
interesting features seen in this model are manifested in this regime.
Finally, the classical conserved angular momentum was kept identical
to the quantum value, $\hbar\sqrt{l^2-1/4}$.

In general, during our numerical calculations, we found that about 80\% of
the $QEE$ were `reliable', in the sense that their imaginary
parts were
$< 10^{-3}$. This is one of the limitations of our numerical determination
of the $QEF$, since being interested in the semiclassical limit
we needed to choose small values of $\hbar$ (indicated in the right hand corner
of Figs. (10-12). This meant that a large number of Fourier
components were needed to accurately represent the $QEF$.
The answers for the $QEF$ were quite reliable when calculating
the Husimi distributions. However, the result were not good enough to do
a RMT statistical analysis of the QFIPO model spectrum.
This is why we resorted to the QFAD model, already discussed in the
first part of this paper. Since we had precisely the opposite difficulty with
the QFAD model we see that  these two representations lead to results
which are complimentary.

The classical analysis was carried out for different values of the angular
momentum $J$. First, we iterated
a single (arbitrarily chosen) initial condition several thousand times,
which typically leads to the chaotic background as shown in the figures.
Imbedded in this background are KAM tori centered around elliptic fixed points,
defined by choosing appropriate initial conditions. In Figure 2(a), we show
several such tori, and in particular, a fixed point of period 2 (marked
as F1 and F2) which was determined earlier analytically. Also shown
in each of the figures is a hyperbolic fixed point of order 6, marked by
its stable and unstable manifolds. The fixed points were determined by
using a modified Powell method \cite{min}
of determining zeros of coupled nonlinear
sets of equations. This method, like all multidimensional root-finding
techniques, requires a good initial guess to converge to the fixed point, but
once given that, determines the root and the Monodromy matrix (the Jacobian
or the determinant of the linearized version of the map equations ) reliably
and
accurately. The fixed point is elliptic, parabolic or hyperbolic if the
discriminant obtained from the eigenvalues ({\it i.e.},
$(Trace)^2-4\cdot (Determinant)$) is negative, zero or positive, respectively.
In all cases, it was verified within numerical error that the map was
area-preserving, {\it i.e.}, the determinant was equal to one. The unstable
manifold was obtained by iterating the map along the direction given by
the eigenvector corresponding to the eigenvalue larger than one. The
{\it stable} manifold is given by the time reversed version of the unstable
one.

Comparison of the Husimi distributions ${\cal F}_{\psi}(\rho_0,p_0)$ with
the classical phase space plots show some striking similarities. There
are, for many $QEF$, many structures which unmistakably
correspond to elliptic, parabolic and hyperbolic periodic orbits, as seen
in Figs. 10(a-c),11(a-c) and 12(a-c).
In the cases of $l=1, 5, 10$, for example, the Husimi representation of one
of the $QEF$ given in Fig. 12(a), 12(b) and 12(c) all sit
on top of the analytic period-two fixed point marked as F.  Also, seen
in the figures there are  Husimis which peak {\it   exactly} on top
of the unstable hyperbolic period-6 fixed point, referred to in the
literature as ``scars''. This correspondence is so robust, in fact, that
often when a good guess to the {\it classical} hyperbolic fixed points
are unavailable, the Husimis are used as a guide to the location of the
fixed point (being unstable, hyperbolic fixed points cannot be located without
a very good initial guess). These enhanced probability densities are
conjectured to play as important a role in quantum mechanics as the
hyperbolic orbits play in classical chaos. Finally, a rare but persistent
occurrence in all the cases considered is that of a single Husimi distributions
peaked simultaneously over {\it both} elliptic and hyperbolic fixed points,
reflecting a purely quantum-mechanical tunneling between the KAM tori.
Here we have only shown representative results of the correspondence between
Husimi distributions and classical solutions.

\section{Discussion and outlook}

In this paper we have presented results of a thorough analysis of the quantum
manifestations of classical chaos in a Fermi accelerating disk (FAD). We
started
by presenting results of the statistical properties of the quasi-energy
spectrum
for both the eigenvalues and the eigenfunctions. Most of the established tests
of the manifestations of chaos in the quantum limit were implemented
successfully with a few exceptions. To wit,
we found clear evidence for the transition from Poissonian to $COE$-like
behavior in the eigenvalue statistics when we changed $\hbar $
from large to small.   We have also connected
the Poisson to $COE$ transition in the eigenvalues to the corresponding
localized to extended regime of the eigenfunctions. In the $COE$ regime the
Porter-Thomas distribution gives a good account of the statistical properties
of the probability amplitudes, while in the Poisson limit the $\chi ^2$ test
fails to give agreement with our results.

While the statistical tests have become the hallmarks of the QMCC it is
in phase space where a direct application
of the correspondence principle can be probed. We found that the standard
description in the FAD representation is not adequate to exhibit
the quantum-classical correspondence. By carrying out a classical Liouville
transformation plus the corresponding quantum unitary transformation, leading
to the Fermi inverted parametric oscillator
(FIPO), the correspondence was implemented explicitly for a selected
set of classical trajectories in terms of their Husimi quesienergy
distributions in phase space.
The classical dynamics was studied in terms of a stroboscopic phase space
plot. Various periodic orbits were found either analytically or numerically
for different values of angular momentum. The quantum quasi-energy
eigenfunctions were represented in phase space using the Husimi distribution.
A direct correspondence was found between the classical stroboscopic
phase space plot and the Husimi distributions of the quantum quasi-energy
eigenfunctions. Periodic orbits present in the classical phase space
were identified with corresponding high probability distribution in the
quantum quasi-energy eigenfunctions. We found the ``scars'' in the
quantum quasi-energy eigenfunctions that correspond to the classical
unstable periodic orbits. We also found
quantum quasi-energy eigenfunctions that corresponds to tunneling
across KAM curves between  stable and unstable periodic orbits.
While classically the KAM curves block the chaotic diffusion, the
quantum tunneling allows it.

Our study extends the analysis carried out in one dimensional driven systems
to quasi-two dimensional ones. This study is then the first step in the
direction of testing the results hereby obtained with experiments, in
particular in mesoscopic quantum dots in the presence of time varying magnetic
fields. We shall discuss this problem elsewhere\cite{bj94}.

\section*{Acknowledgements}

We thank G. Farmelo and O. Bohigas for contributions and discussions
in the initial stages of this research, and F. Leyvraz and T. Seligman for
useful discussions on eigenvector statistics.
This work has been supported by the Office of Naval Research grant number
ONR-N00014-92-1666.


\appendix
\section*{}

In this Appendix, we set up the matrix elements of the evolution
operator $U$. The Fourier sine basis we choose is
\begin{equation}
\langle\rho,\phi|l,m\rangle =
\sqrt{{2\over \pi}}\,\sin(m\,\pi\,\rho) \cos(l\,\phi)
\end{equation}
The basis is orthonormal :
\begin{equation}
\langle l',m|l,n\rangle = \delta_{l',l}
\delta_{m,n}
\end{equation}
Then, the matrix elements of $H_0$ are
\begin{equation}
\langle l',m|H_0|l,n\rangle = H_0^{mn}
\delta_{l',l}
\end{equation}
where
\begin{eqnarray}
H_0^{mn} &= -{\strut\displaystyle{\hbar^2}\over\displaystyle{2}}\langle
m|\,(p^2 + \epsilon^2 \rho^2)
|n\rangle + {\strut\displaystyle{(l^2 - {1\over 4})\hbar^2}\over
\displaystyle{2}}\langle m|
{\strut\displaystyle{1}\over\displaystyle{\rho^2}} |n\rangle \\
&= {\strut\displaystyle{\hbar^2\pi^2 m^2}\over\displaystyle{2}}\,\delta_{m,n}
- {\strut\displaystyle{\epsilon^2}\over\displaystyle{\pi^2}}\cdot
\left\{ \begin{array}{ll}
(-1)^{m-n}{\strut\displaystyle{4mn}\over\displaystyle{(m^2-n^2)^2}} & m\not=n
\\
{\strut\displaystyle{\pi^2}\over\displaystyle{6}} -
{\strut\displaystyle{1}\over\displaystyle{4m^2\pi^2}} & m=n \end{array}
\right. \\
&+ {\strut\displaystyle{(l^2-{1\over 4})\hbar^2}\over\displaystyle{2}}\pi
\left\{ (m+n){\rm Si}\left[\pi(m+n)\right] - (m-n)
{\rm Si}\left[\pi(m-n)\right] \right\}
\end{eqnarray}
where Si($x$) is the sine integral,
\begin{equation}
{\rm Si}(x) = \int_0^x {\sin(u)\over u}\,du .
\end{equation}
We diagonalized $H_0$ numerically to obtain its eigenvalues
and eigenvectors :
\begin{eqnarray}
H_0|{\mu}\rangle &= \hbar\,\omega_{\mu}|{\mu}\rangle \\
|{\mu}\rangle &= \sum_{n=1}^N \langle n|\mu\rangle |{n}\rangle \\
&\equiv \sum_{n=1}^N c_{n,\mu} |{n}\rangle .
\end{eqnarray}
Here we dropped the angular momentum quantum number index $l$,
because angular momentum conservation forbids transitions between states
of different $l$.
Using the completeness and orthonormality properties of the basis set,
we obtain the matrix elements of $U$ as
\begin{eqnarray}
\langle m|U|n\rangle = \sum_{\mu,\nu=1}^N e^{-i(\omega_{\mu}+
\omega_{\nu})/2} \times \\
&\sum_{p,q,r}^{\,} c_{n,\mu}^*\,c_{p,\mu}\,c_{q,\nu}^*\,c_{r,\nu}
\langle m|\,e^{i\epsilon\hat\rho^2/\hbar}\,|r\rangle\langle q|\,e^{-i\epsilon
\hat\rho^2/\hbar}\,|p\rangle .
\end{eqnarray}
The last two quantities are again evaluated numerically by using methods
for calculating integrals with rapidly oscillatory integrands.

%
%

%
%
%
%
\begin{figure}
\caption{ Poincar{\' e} surfaces of section in the
$(r,p_r)$ plane.
(a) Almost regular behavior for small wall amplitude ($\epsilon=0.001,l=1)$.
(b) Intermediate case where regular and chaotic motions
coexist ($\epsilon=1.0,l=10$). (c) Fully chaotic
case ($\epsilon=10, l=100$).}
\label{fig1}
\end{figure}

\begin{figure}
\caption{ (a)-(d) Transition from Poisson to Wigner
for $\Pi _{\alpha}$ as the tuning parameter
is changed. The solid line in (a-d) corresponds to a $\Pi _{\alpha}(s)$ fit
with the $\alpha$ value obtained from the best straight line
fit to $\ell n \ell n \left[1-\Pi_{\alpha}(s)\right]^{-1}$.
The Brody fit is essentially on top of the calculated data.
The values of $(\hbar, l)$ were (a) (1.0, 10),
(b) (0.5, 20),  (c) (0.1, 100) and (d) (0.01, 1000) for $\epsilon=1$.
The repulsion parameter $\alpha$ for (a-d) are $0.10, 0.20,
0.60$ and $0.86$, respectively. }
\label{fig2}
\end{figure}

\begin{figure}
\caption{(a)-(d) Nearest-neighbor spacing distribution $P(s)$
for the same parameters as in Fig. 2. The dashed line in (a-b)  corresponds to
the Poisson limit, and to the $COE$ case in (c) and (d). The solid line in
(a-d) correspond to a $P_{\alpha}$ fit with the $\alpha$ values obtained in
the fits of fig. 2.}
\label{fig3}
\end{figure}

\begin{figure}
\caption{(a)-(d) Fits of the data to the Izrailev
formula $P_{\beta}(s)$ (solid line). The dots are the fits as given
by $\beta_{S}$ (see text for details). The best-fit and predicted values
for $(\beta_{hist}, \beta_{S})$ are,
(a) (0.20,0.05), (b) (0.22,0.12), (c)
(0.48,0.50) and (d) (1.14,1.00). }
\label{fig4}
\end{figure}

\begin{figure}
\caption{The $<\Delta_3(L)>$ (upper plot) and the
$\Sigma^2(L)$ statistics (lower plot). The parameter values are the same as in
Fig. 2 with (a) $+$, (b) $\square $ , (c) $\diamond $, and (d) for $\times$.
The dashed line is the exact Poisson result while the continuous line is the
corresponding $COE$ RMT results. We observe a continuous transition from
Poisson
to $COE$. In the Poisson case the calculated values are slightly above the
predicted behavior but it does not show the bending characteristic of picket
fence  spectra. In the $COE$ regime no knee is found either.}
\label{fig5}
\end{figure}

\begin{figure}
\caption{(a)-(d) Distribution of amplitudes of $QEF$
with the natural basis states for the same parameter values  as in Fig. 2.
Close to the $COE$ limit (d) the amplitudes are nearly gaussian or
Porter-Thomas randomly distributed. Away from this limit the distributions
are not well fitted by the $\chi^2$ distribution with a significant difference
seen close to the Poisson limit. This discrepancy is explained in the text.
The values of $\nu$ from the fits are, (a) $0.10$, (b) $0.15$, (c) $0.52$,
and (d) $1$.}
\label{fig6}
\end{figure}

\begin{figure}
\caption{(a)-(d) Logarithm of the pair-correlation function
defined in Eq.(64) as a function of $\Delta $ for the same parameter values
as in Fig.2. The transition from localized to extended behavior, exemplified
by the value of the localization length $\xi$ as compared to the matrix sizes
(N=400 in this case).}
\label{fig7}
\end{figure}

\begin{figure}
\caption{ Classical Poincar\'e section for $l=1$ (a), $5$ (b) and $10$
(c). The period-2 fixed points are denoted as F.
The  hyperbolic fixed points in (b) are obtained as explained in the
text. Here, and in the following, $\hbar=0.026, \epsilon=1.0$}
\label{fig8}
\end{figure}

\begin{figure}
\caption{Schematic representation of the trajectory of the particular
analytically calculated period-2 orbit, as explained in the text.
The vertical scale is in
arbitrary energy units, and the small separations correspond to the same
energy. The horizontal scale is in arbitrary length units.}
\label{fig9}
\end{figure}

\begin{figure}
\caption{ (a-c) Contour plots of the Husimi distribution of a
$QEF$ which corresponds to the three classical solutions
corresponding
to Fig. 8(a) ($l=1$). In (a) we have to Husimi distribution that corresponds
to the period-2 solution calculated analytically in the classical limit.
(b) Husimi distribution of another $QEF$ which corresponds to
the period-6 hyperbolic
orbit marked by its stable and unstable manifolds - a `scarred' eigenfunction.
(c) A mixed case, where a {\it single} $QEF$ corresponds to both
a stable,
period-2 elliptic orbit, and the unstable, hyperbolic period-6 orbit.}
\label{fig10}
\end{figure}

\begin{figure}
\caption{ Same as in the previous figure corresponding to the case $l=5$
of Fig. 8(b).
}
\label{fig12}
\end{figure}

\end{document}